\documentclass[%
 reprint,
 amsmath,amssymb,
 aps,
]{revtex4-1}

\usepackage{graphicx}
\usepackage{dcolumn}
\usepackage{bm}

\usepackage{definition}

\begin{document}

\preprint{IFT/12/03}

\title{Neutrino oscillations in the formal theory of scattering}

\author{Stanis\l{}aw D. G\l{}azek}
\author{Arkadiusz P. Trawi\'nski}
\affiliation{
Institute of Theoretical Physics, Faculty of Physics, University of Warsaw 
}

\date{\today}

\begin{abstract}
	Scattering theory in the Gell-Mann and Goldberger formulation 
is slightly extended to render a Hamiltonian quantum mechanical 
description of the neutrino oscillations.

\end{abstract}


\maketitle

\section{\label{sec:introduction} Introduction}

The concept of neutrino oscillation has become 
very familiar to physicists during recent years 
and the subject is discussed in numerous articles.
Nevertheless, there does not appear to be a 
Hamiltonian treatment which proceeds from fundamental
quantum-mechanical principles. The present work 
represents an attempt to fill this deficit starting
from the Gell-Mann and Goldberger formulation 
of scattering theory~\cite{GellMann:1953zz}. There 
are very few new results, but the interpretation is 
somewhat new. The theory is limited to scattering 
in vacuum and does not include matter 
effects~\cite{Wolfenstein:1977ue,Mikheev:1986gs}.

The neutrino oscillations have been identified 
as a potential source of information about 
fundamental aspects of particle theory by 
Pontecorvo~\cite{Pontecorvo:1967fh}. Bilenky 
and Pontecorvo provided an initial theoretical 
analysis of the neutrino oscillation as a 
subject of its own merits~\cite {Bilenky:1977ne}.
Kayser~\cite{Kayser:1981ye} introduced wave 
packets for neutrinos. Rich~\cite{Rich:1993wu}
offered a space-time approach where neutrino
emission, propagation, and absorption are 
treated as a single process. Giunti, Kim, Lee, 
and Lee~\cite{Giunti:1993se} further developed 
an approach based on Feynman diagrams and wave 
packets. Recently, Akhmedov and Kopp~\cite{Akhmedov:2010ms}
summarized the current theory status.

Davis, Harmer and Hoffman reported a deficit 
in the flux of neutrinos from the sun in 1968 
\cite{Davis:1968cp}. Major collaboration reports 
on measurements of the neutrino oscillation are
available from Super-Kamiokande~\cite{Fukuda:1998mi}, 
SNO~\cite{Ahmad:2002jz}, and KamLAND~\cite{Eguchi:2002dm}. 
Most recent results of T2K and MINOS are described 
in Refs.~\cite{Abe:2012gx} and~\cite{Adamson:2012rm}, 
respectively.

Section \ref{sec:amplitude} introduces the 
required elements of the formal scaterring 
theory. The neutrino oscillations are discussed 
in Section \ref{sec:NO}, where it is explained 
how the standard formula emerges in their theory. 
Section \ref{sec:summary} concludes the paper. 
Appendix \ref{sec:appendix} explains relevant 
details of perturbative energy denominators.

\section{\label{sec:amplitude} Formal theory}

The formal scattering theory of Gell-Mann 
and Goldberger~\cite{GellMann:1953zz} starts
form a full Hamiltonian $H$ which is a sum 
of a free part $H_0$ and the interaction part 
$H_I$. An incoming state $\ket{\Psi_i(t)}$ of 
energy $E_i$ is gradually built over time 
$\tau = \epsilon^{-1}$ according to the formula
\beq
\ket{\Psi_i(t)}
\es 
\epsilon \int^0_{-\infty} dT \, e^{\epsilon T} \, 
e^{-iH(t-T)}\ket{\Phi_i(T)}\\
\es
e^{-iHt}\frac{\epsilon}{\epsilon+i(H-E_i)}\ket{\phi_i}\, .
\eeq
In this formula, the state $\ket{\phi_i}$  
is an eigenstate of $H_0$,
\beq
H_0 \ket{\phi_i} \es E_i \ket{\phi_i} \, ,
\eeq
so that
\beq
\ket{\Phi_i(T)} \es e^{-iE_iT}\ket{\phi_i} \, .
\eeq
The norms of states are not changing 
with time. The probability that the 
system is in a state $|\Phi_f\rangle$ 
at time $t$ is
\beq
\omega_{fi}(t)
\es 
\frac{ |A_{fi}(t)|^2}{||\Phi_f||^2 ||\Psi_i||^2},
\eeq
where the amplitude $A_{fi}(t)$ is 
\beq
	\label{At}
	A_{fi}(t)
	\es
	\bra{\phi_{f}}
	\frac{i\epsilon \, e^{i(E_f-H)t}}{E_i-H+i\epsilon}
	\ket{\phi_{i}}
	\label{eq:A(t)} \, .
\eeq
The differential cross section 
for the transition $i \rightarrow f$ is 
equal to the transition rate, 
\beq
P_{fi}(t)
\es 
{ d \over dt} \, \omega_{fi}(t) \, ,
\eeq
divided by the flux of incoming particles.

Gell-Mann and Goldberger argue that for $t \ll \tau$ 
\beq
    \dot {|A_{fi}(t)|^2} 
    & \approx & 
    \dot {|A_{fi}(0)|^2} 
\eeq
and they focus on the case of $t=0$. However,  
the time $t$ in the neutrino oscillation 
experiments is typically greater than $\tau$. 
For example, in the T2K experiment, the initial 
pion states are generated from a carbon target 
of size of about 1 m (using energetic protons)
and the pions only move about 100 m before 
$H_I$ nearly certainly turns them into states 
of neutrinos and muons. This means that $\tau$ 
is shorter than about $ \sim 1$ $\mu$s. Since 
the initial neutron is located about 300 km 
away, the time $t$ must be greater than 300 km$/c 
\sim$ 1 ms. The time $t$ is thus expected to be 
more than $10^3$ times greater than $\tau$.

In addition, the formal scattering theory 
requires that the linear dimension of the 
region in which the states are normalized 
is much greater than $\tau$ times the group 
velicity of the incident particle wave trains. 
This condition is satisfied in the T2K 
experiment when the required region contains 
both Tokai and Kamioka.

To describe scattering in these circumstances, 
the formal scattering formula needs to be 
extended to large times $t$. Using notation
\beq
\label{delta}
	\delta^*_{fi}(t)
	\es
	{{i\epsilon}\over{E_i-E_f+i\epsilon}} 
	\bra{\phi_{f}} \, e^{i(E_f-H)t}\ket{\phi_{i}},\\
\label{R}
	R_{fi}(t, \epsilon)
        \es \bra{\phi_f} \,H_I e^{i(E_f-H)t} 
        \frac{i\epsilon }{E_i -H + i\epsilon}\ket{\phi_i} \\
\es \bra{\Phi_f(t)}H_I\ket{\Psi_i(t)} \, , 
\eeq
one obtains from Eq. (\ref{At}) that
\beq
	\label{eq:A+}
	A_{fi}(t) \es \delta^*_{fi}(t) + 
        \frac1{E_i-E_f+i\epsilon} R_{fi}(t, \epsilon),\\
	\dot A_{fi}(t) \es  -iR_{fi}(t, \epsilon)  \, .
\eeq
In the spirit of~\cite{GellMann:1953zz}, 
one can suggest that Eq. (\ref{eq:A+}) is 
useful because it exhibits the energy 
dependence of $A_{fi}(t)$ in the vicinity
of $E_f = E_i$. The resulting transition 
rate is
\beq
\label{eq:dotAA*}
	\dot{|A_{fi}(t)|^2}
	\es  
	2 \, \text{Im}\left[\delta_{fi}(t) \, 
        R_{fi}(t,\epsilon)\right]
	\np
	\frac{2\epsilon}{(E_i-E_f)^2+\epsilon^2}
        |R_{fi}(t,\epsilon)|^2\, .
\eeq
Since in the neutrino oscillation experiments
such as T2K the time $\tau$ is shorter than $t$,
one cannot send $\epsilon = \tau^{-1}$ to 0 for
finite $t$. Instead, $\epsilon$ approximately 
accounts for the experimental energy uncertainty 
and determines the width of density in energy of 
final states $f$ at energy $E_i$. 

The parameter $\epsilon$ also smoothes out all 
contributing amplitudes as functions of energy. 
As a result of this smooting, the interference 
effects in the complete transition rate are 
described by the standard oscillation formula, 
see Eqs.~(\ref{P}) and~(\ref{result}). In the 
remaining part of the article, it is shown how 
this happens when one approximates the full 
Hamiltonian $H$ in the exponential factors in 
Eqs.~(\ref{delta}) and~(\ref{R}) by $H_0$. 

Further analysis will involve specification 
of $H_I$. We choose to continue focusing on 
the example of T2K.

\section{\label{sec:NO} Neutrino oscillation}

In the T2K experiment, neutrino
oscillation occurs in the process 
\beq
	\label{process}
	\pi^+ n 
	& \rightarrow & 
	\mu^+ \, \mu^- \, p \, ,
\eeq
where the initial state consists of a 
$\pi^+$-meson prepared in Tokai and 
a neutron, $n$, located in Kamioka.
The final $\mu^+$ is produced in decay 
of $\pi^+$ in Tokai, and $\mu^-$ and 
proton, $p$, emerge in Kamioka. The 
distance covered by the intermediate 
neutrinos, which are created in the 
interaction responsible for $\pi^+$-decay 
and annihilated in the interaction that 
changes $n$ to $p$ and creates $\mu^-$, 
is practically equal to the distance 
from Tokai to Kamioka. We assume that 
the only particle detected in the final 
state is $\mu^-$. Detection of other 
particles does not change our conclusions.

Since the dominant interaction processes
are $\pi^+ \rightarrow \mu^+ \nu_\mu$ and 
$n \nu_\mu \rightarrow p \mu^-$ and the 
neutrinos carry momenta and energies on 
the order of 1 GeV, the most appropriate 
interaction Hamiltonian density for 
obtaining $H_I$ by integration over space
is~\cite{Feynman:1958ty,GellMann:1960np,
Weinberg:1991um}
\beq
\label{eq:cHI}
	\cH_I
	\es \frac{G_F}{\sqrt2}\cos\vartheta_C \,\, 
        \overline \mu\gamma^\alpha(1-\gamma_5)\nu_\mu \,\,
	 \overline p\gamma_\alpha(1-g_A\gamma_5)n
	\nm i\frac{F_\pi}{\sqrt2} \, 
        \overline \nu_\mu\gamma^\alpha(1-\gamma_5)\mu
	\,\, \partial_\alpha\pi^\dagger
		+ h.c. \quad .
\eeq
The resulting $H_I$ is translation invariant 
and hence conserves three-momentum. The 
interaction does not conserve the eigenvalues 
of $H_0$, which are sums of energies of the form
$E = \sqrt{m^2 + \vec p^{\, 2} }$ for any particle 
of mass $m$ and three-momentum $\vec p$.

The central issue is that the neutrino states
that are created or annihilated by $H_I$ are
not eigenstates of $H_0$. One can consider 
$H_I$ at the hadronic level of Eq.~\eqref{eq:cHI}, 
or at the level of electroweak bosons that 
interact with quarks in $\pi^+$ and nucleons. 
In both cases, the neutrino states that are 
created or annihilated in the interactions with 
muons, called $\mu$-neutrinos and denoted by 
$\nu_\mu$, are thought to be properly described 
by a unique combination of (most likely) three 
quantum fields $\nu_i(x)$, $i = 1, 2, 3$, each 
of which corresponds to neutrino states that 
are eigenstate of $H_0$ and having different masses. 
In the case of $\nu_\mu$, 
\beq
	\nu_\mu(x) \es \sum_{i=1}^{3} U_{\mu i} \, \nu_i(x)\, .
\eeq
Recent data on the mixing coefficients, 
$U_{\mu i}$, can be found in~\cite{Nakamura:2010zzi}.

The standard formula for neutrino oscillation 
in the quantum process~\eqref{process} is 
obtained from Eq. (\ref{eq:dotAA*}) when the 
Hamiltonian $H$ in the exponential factors in Eqs. 
(\ref{delta}) and (\ref{R}) is approximated by 
$H_0$. In addition, one also expands $(E_i -H 
+i\epsilon)^{-1}$ in Eq. (\ref{R}) up to second 
power of $H_I$. The result is
\beq
\label{RPT}
R_{fi}(t,\epsilon)
	\es \bra{\phi_f}
	H_I \frac{e^{ i(E_f-H_0)t}}{E_i - H_0 + i\epsilon}H_I
	\ket{\phi_i}+\cdots \, ,
\eeq
where $\ket{\phi_i} \rs \ket{\pi^+ n}$ and
$\ket{\phi_f} \rs \ket{\mu^+ \, \mu^- \, p }$.

There are only two types of intermediate 
states that contribute to $R_{fi}$ between 
the operators $H_I$ in Eq.~(\ref{RPT}). The 
first type contains a neutrino, $\mu^+$, 
and $n$. The second type contains an anti-neutrino, 
$\pi^+$, $\mu^-$, and $p$. Only the first type 
of intermediate states can lead to a small 
energy difference in denominator, due to 
$E_i-H_0+i\epsilon$. Thus, the transition 
rate in Eq.~(\ref{eq:dotAA*}) is dominated 
by the contribution from the first type of 
intermediate states and the second-type 
contribution can be neglected.

Namely, the energy-denominator associated 
with an intermediate state with a neutrino 
of mass $m_i$ is 
\beq
\label{Di}
D_i \es E_\nu - E_{\nu_i} + i \epsilon \, , 
\eeq 
where $E_\nu$ denotes the energy transfer 
from $\pi^+$ to $n$, 
\beq
\label{Enu}
E_\nu \es E_{\pi^+} - E_{\mu^+} \, ,
\eeq 
and the neutrino energy eigenvalue 
of $H_0$ is 
\beq
\label{H0eigenvalue}
E_{\nu_i} \es \sqrt{m_i^2 + \v p_\nu^2 } \, .
\eeq
All eigenvalues ascribed by $H_0$ to all 
intermediate states with neutrinos
contain $E_{\nu_i}$ with one and the same 
physical momentum transfer $\vec p_\nu$ 
irrespective of the neutrino mass. The 
value of $\vec p_\nu$ is implied by the 
three-momentum conservation in the 
translation-invariant Hamiltonian 
interaction terms.

While the real part of $D_i$ in Eq.~(\ref{Di}) 
must be on the order of $\epsilon$ or smaller 
according to the density in energy exhibited 
in Eq.~(\ref{eq:dotAA*}), the energy denominators 
associated with the intermediate states with 
anti-neutrinos have real parts on the order 
of $- 2 E_\nu$. Therefore, the second-type 
contributions to the transition rate are about 
$(\epsilon/E_\nu)^2$ times smaller than the 
first-type, and are neglected.

Consequently, $R_{fi}$ in Eq.~(\ref{RPT}) is
a sum of three amplitudes each of which 
corresponds to the intermediate state with 
a virtual neutrino of a different mass. 
When one evaluates the time derivative 
of the modulus squared of the sum at time 
$t = L/c$, one obtains 9 terms which together 
combine to 
\beq
	\label{PL}
	&&\left.{d \over dt} {|A_{fi}(t)|^2}\right|_{t=L} \nn
	\es
(2\pi)^3\delta^{(3)} ( \v p_\pi + \v p_n 
                    - \v p_{\mu^+} -\v p_p -\v p_{\mu^-}) \\
	& \times &
        \frac{2\epsilon}{\left( E_\pi+E_n
                              - E_{\mu^+}-E_{\mu^-}-E_p\right)^2
                        +\epsilon^2}
        \, |F_\nu|^2  \, G_\nu \nonumber
\eeq
in a self-explanatory fashion. The factor 
$|F_\nu|^2$ results from $H_I$. The factor 
$G_\nu$ is
\beq
\label{G}
G_\nu \es \sum_{i,j}  |U_{\mu j} U_{\mu i}|^2  e^{i(E_{\nu_j}-E_{\nu_i})L} \,
G_{ji} \, ,
\eeq
where 
\beq
\label{Dji}
G_{ji} \es 
{ 1 \over 4E_{\nu_j}E_{\nu_i}} \, 
\frac{1}{D_j^*}
\frac{1}{D_i} \,
 \, .
\eeq
The energy denominators $D_i$ are different
for different values of $i$ if the masses
$m_i$ in the eigenvalues $E_{\nu_i}$ are 
different. The differences between the eigenvalues
are very small if the mass differences are very 
small and if the masses are very small in 
comparison to the physical momentum transfer 
$|\vec p_\nu|$. Monte-Carlo studies of T2K 
Collaboration suggest that these conditions 
are satisfied.

Under the condition that
\beq
\label{condition}
{|m_j^2 \pm m_i^2| \over 2 | \v p_\nu| } 
& \ll & \epsilon \, ,
\eeq
one obtains the approximate $G_{ji}$ that
does not depend on the neutrino masses.
This is shown in Eqs.~(\ref{eq:approx1}) 
and~(\ref{eq:approx2}) in Appendix~\ref{sec:appendix}. 
Taking the common approximate factor $G_{ji}$ out
of the sum of 9 terms, one arrives at
\beq
\label{G23}
  G_\nu \es { 1 \over 4 \v p_\nu^2} \,
       	   { \sum_{i,j}  |U_{\mu j} U_{\mu i}|^2 \, 
           e^{i(m^2_j-m^2_i)\, L/(2|\v p_\nu|)}
           \over
           \left(E_\nu - |\v p_\nu| \right)^2+\epsilon^2 }\, .
\eeq
This result implies that the rate of counting 
$\mu^-$ in a distant detector depends only on
the ratio $L/|\v p_\nu|$ as in the standard
neutrino oscillation formula.

Namely, the numerator in $G_\nu$ in Eq. 
(\ref{G23}) is the standard, distance-dependent 
$\mu^-$-detection probability,
\beq
	P_{\mu \to \mu}(L)
	\es
	\sum_{i,j} |U_{\mu j} U_{\mu i}|^2
	e^{i(m^2_j-m^2_i)L/(2|\v p_\nu|) } \\
	& \approx &
	1 - \sin ^2 \left(2\theta_{23} \right)
	\sin^2 { \Delta m_{23}^2 \, L \over 4|\v p_\nu| } \, .
\label{P}
\eeq
More precisely, the ratio of $\mu^-$ 
counting-rates at two different distances 
between the neutrino detector and a 
$\pi^+$-source, such as $L_{far} \approx 300$ 
km and $L_{near} \approx 280$ m in the T2K 
experiment, is
\beq
\label{result}
        {{\left.{d \over dt} {|A_{fi}(t)|^2}\right|_{t=L_{far}}}
	\over
	{\left.{d \over dt} {|A_{fi}(t)|^2}\right|_{t=L_{near}}}}
	\es
	{{P_{\mu \to \mu}(L_{far})}\over{P_{\mu \to \mu}(L_{near})}} \, .
\eeq
This is how the perturbative scattering theory 
explains the standard neutrino oscillation formula.

\section{\label{sec:summary} Conclusion}

A simple extension of the Gell-Mann--Goldberger 
formulation of scattering theory to the case of 
a long base-line experiment, leads to the standard 
neutrino oscillation formula that describes 
the distance-dependent ratio of muon-counting 
rates via Eq. (\ref{result}). This conclusion 
holds provided the condition (\ref{condition}) 
for a pion-beam preparation time $\tau = 
\epsilon^{-1}$ is satisfied and one neglects 
all interaction effects between the initial 
pion decay and neutrino absorption in a 
detector. This is formally facilitated 
by replacing $H$ with $H_0$ in the exponents
in Eqs.~(\ref{delta}) and~(\ref{R}). 

The oscillation formula is a result of the 
interference between scattering amplitudes
mediated by virtual states with neutrinos
of different masses. All these states mediate 
a transfer of the same physical three-momentum 
$\v p_\nu$ and energy $E_\nu$. However, their 
contributions to the total amplitude differ 
as functions of $\v p_\nu$ and $E_\nu$. According 
to Eq.~(\ref{G}), only a sufficiently large 
$\epsilon$ can smooth out the differences between 
factors $G_{ji}$ and yield Eq.~(\ref{G23}), so 
that the net effect of the interference can 
be described using the standard oscillation 
formula.

The condition (\ref{condition}) indicates 
that the time of preparing the beam of $\pi^+$ 
mesons must be sufficiently short, the neutrino 
momentum must be sufficiently large, and the 
neutrino masses sufficiently small for the 
oscillation to occur in agreement with the 
standard formula. Otherwise, one might expect 
deviations, whose details can be deduced from 
Eq.~(\ref{G}).

Regarding the replacement of $H$ by $H_0$ in 
the exponents in Eqs. (\ref{delta}) and 
(\ref{R}), one may observe that precise 
calculations of weak interaction effects
following from $\cH_I$ of Eq. (\ref{eq:cHI}), 
require understanding of the cutoff dependence 
that appeared already in Pontecorvo's 
work~\cite{Pontecorvo:1967fh}. One also ought 
to consider interactions with 
matter~\cite{Wolfenstein:1977ue,Mikheev:1986gs}.

It should be stressed that the conclusion 
concerning the ratio of transition rates does
not automatically translate to the ratios of
entire cross sections. Evaluation of cross sections 
includes averaging over incoming and integration 
over outgoing states and may involve wave packets, 
density matrices, entanglement, and various 
experimental cuts. Such evaluation is a formidable 
task in a fundamental theory. For example, if 
neutrinos are coupled through intermediate bosons 
to quarks, and one attempts to evaluate the complete 
scattering process including the quark structure of 
hadrons, in addition to the nuclear binding effects, 
an entire host of additional theoretical issues 
arise. To be more specific, the Fermi motion of 
quarks in nucleons, or of nucleons in nuclei, is 
a subject of study in its own right \cite{Glazek:1987cd} 
and currently available level of theoretical analysis 
could certainly be improved, e.g., see Eq. (2) in 
\cite{Abe:2012gx}.

Finally, we note that the Gell-Mann and Goldberger 
discussion of a connection between the formal 
scattering theory and $S$-matrix formalism in the 
interaction picture involves the limit of $\epsilon 
\rightarrow 0$. This limit requires that the beam
preparation time $\tau$ is much longer than the 
period in which interactions causing scattering 
may happen. If this condition is not satisifed and 
one cannot take the limit of $\epsilon \rightarrow 0$, 
one has to proceed from fundamental quantum-mechanical 
principles rather than taking advantage of simplified 
formulae in which $\epsilon \rightarrow 0$.

\newpage
{\tiny .}\vskip-.4in


\appendix
\section{\label{sec:appendix} Energy denominators }

Factor $G_{ji}$ in Eq.~(\ref{Dji}) can be rewritten as
\beq
	\label{eq:approx1}
	G_{ji} 
        \es 
        { 1 \over 4 E_{\nu_j} E_{\nu_i}}
	\frac1{\left(E_\nu - \frac{E_{\nu_j}+E_{\nu_i}}{2}\right)^2
	+ \left(i \frac{E_{\nu_j}-E_{\nu_i}}{2} + \epsilon\right)^2}\nn
	& \approx & { 1 \over 4 E_{\nu_j} E_{\nu_i}}
	\frac1{\left(E_\nu -\frac{E_{\nu_j}+E_{\nu_i}}{2} \right)^2
        +\epsilon^2}\, ,
\eeq
which is a valid approximation when $|E_{\nu_j}-E_{\nu_i}|/2 
\ll \epsilon$. In addition, using condition (\ref{condition}), one obtains 
\beq	
        \label{eq:approx2}
        G_{ji} & \approx & { 1 \over 4 \v p_\nu^2}
	\frac1{\left(E_\nu - |\v p_\nu| \right)^2+\epsilon^2}\, ,
\eeq
which does not depend on the neutrino masses and leads
to Eq.~(\ref{G23}).

Eq. (\ref{eq:approx1}) differs from Eq. (\ref{eq:approx2})
by the terms that in a leading approximation are inversely 
proportional to the neutrino momentum and directly proportional 
to the heaviest neutrino mass squared times $2 \Delta/(\Delta^2 
+ \epsilon^2)$, where $\Delta = E_\nu - |\v p_\nu|$. Therefore, 
the leading correction to the standard oscillation formula 
due to using Eq.~(\ref{G23}) instead of Eq.~(\ref{G}), may be 
expected to be smaller than $c\tau/L$, which currently means 
smaller than $10^{-3}$ in T2K. In any case, its actual size 
depends on the pion beam preparation and the Monte Carlo simulations 
may even average it to nearly 0 due to the varying sign of $\Delta$ 
in data sampling. A special sampling that secures only inclusion of 
cases with one sign of $\Delta$ would be required to facilitate 
studies of this correction.


\bibliography{bibliography}

\begin{thebibliography}{20}%
\makeatletter
\providecommand \@ifxundefined [1]{%
 \@ifx{#1\undefined}
}%
\providecommand \@ifnum [1]{%
 \ifnum #1\expandafter \@firstoftwo
 \else \expandafter \@secondoftwo
 \fi
}%
\providecommand \@ifx [1]{%
 \ifx #1\expandafter \@firstoftwo
 \else \expandafter \@secondoftwo
 \fi
}%
\providecommand \natexlab [1]{#1}%
\providecommand \enquote  [1]{``#1''}%
\providecommand \bibnamefont  [1]{#1}%
\providecommand \bibfnamefont [1]{#1}%
\providecommand \citenamefont [1]{#1}%
\providecommand \href@noop [0]{\@secondoftwo}%
\providecommand \href [0]{\begingroup \@sanitize@url \@href}%
\providecommand \@href[1]{\@@startlink{#1}\@@href}%
\providecommand \@@href[1]{\endgroup#1\@@endlink}%
\providecommand \@sanitize@url [0]{\catcode `\\12\catcode `\$12\catcode
  `\&12\catcode `\#12\catcode `\^12\catcode `\_12\catcode `\%12\relax}%
\providecommand \@@startlink[1]{}%
\providecommand \@@endlink[0]{}%
\providecommand \url  [0]{\begingroup\@sanitize@url \@url }%
\providecommand \@url [1]{\endgroup\@href {#1}{\urlprefix }}%
\providecommand \urlprefix  [0]{URL }%
\providecommand \Eprint [0]{\href }%
\providecommand \doibase [0]{http://dx.doi.org/}%
\providecommand \selectlanguage [0]{\@gobble}%
\providecommand \bibinfo  [0]{\@secondoftwo}%
\providecommand \bibfield  [0]{\@secondoftwo}%
\providecommand \translation [1]{[#1]}%
\providecommand \BibitemOpen [0]{}%
\providecommand \bibitemStop [0]{}%
\providecommand \bibitemNoStop [0]{.\EOS\space}%
\providecommand \EOS [0]{\spacefactor3000\relax}%
\providecommand \BibitemShut  [1]{\csname bibitem#1\endcsname}%
\let\auto@bib@innerbib\@empty
\bibitem [{\citenamefont {Gell-Mann}\ and\ \citenamefont
  {Goldberger}(1953)}]{GellMann:1953zz}%
  \BibitemOpen
  \bibfield  {author} {\bibinfo {author} {\bibfnamefont {M.}~\bibnamefont
  {Gell-Mann}}\ and\ \bibinfo {author} {\bibfnamefont {M.~L.}\ \bibnamefont
  {Goldberger}},\ }\href {\doibase 10.1103/PhysRev9.1.398} {\bibfield
  {journal} {\bibinfo  {journal} {Phys. Rev.}\ }\textbf {\bibinfo {volume}
  {91}},\ \bibinfo {pages} {398} (\bibinfo {year} {1953})}\BibitemShut
  {NoStop}%
\bibitem [{\citenamefont {Wolfenstein}(1978)}]{Wolfenstein:1977ue}%
  \BibitemOpen
  \bibfield  {author} {\bibinfo {author} {\bibfnamefont {L.}~\bibnamefont
  {Wolfenstein}},\ }\href {\doibase 10.1103/PhysRevD.17.2369} {\bibfield
  {journal} {\bibinfo  {journal} {Phys.Rev.}\ }\textbf {\bibinfo {volume}
  {D17}},\ \bibinfo {pages} {2369} (\bibinfo {year} {1978})}\BibitemShut
  {NoStop}%
\bibitem [{\citenamefont {Mikheev}\ and\ \citenamefont
  {Smirnov}(1985)}]{Mikheev:1986gs}%
  \BibitemOpen
  \bibfield  {author} {\bibinfo {author} {\bibfnamefont {S.}~\bibnamefont
  {Mikheev}}\ and\ \bibinfo {author} {\bibfnamefont {A.}~\bibnamefont
  {Smirnov}},\ }\href@noop {} {\bibfield  {journal} {\bibinfo  {journal}
  {Sov.J.Nucl.Phys.}\ }\textbf {\bibinfo {volume} {42}},\ \bibinfo {pages}
  {913} (\bibinfo {year} {1985})}\BibitemShut {NoStop}%
\bibitem [{\citenamefont {Pontecorvo}(1968)}]{Pontecorvo:1967fh}%
  \BibitemOpen
  \bibfield  {author} {\bibinfo {author} {\bibfnamefont {B.}~\bibnamefont
  {Pontecorvo}},\ }\href@noop {} {\bibfield  {journal} {\bibinfo  {journal}
  {Sov.Phys.JETP}\ }\textbf {\bibinfo {volume} {26}},\ \bibinfo {pages} {984}
  (\bibinfo {year} {1968})}\BibitemShut {NoStop}%
\bibitem [{\citenamefont {Bilenky}\ and\ \citenamefont
  {Pontecorvo}(1977)}]{Bilenky:1977ne}%
  \BibitemOpen
  \bibfield  {author} {\bibinfo {author} {\bibfnamefont {S.~M.}\ \bibnamefont
  {Bilenky}}\ and\ \bibinfo {author} {\bibfnamefont {B.}~\bibnamefont
  {Pontecorvo}},\ }\href@noop {} {\bibfield  {journal} {\bibinfo  {journal}
  {Comments Nucl. Part. Phys.}\ }\textbf {\bibinfo {volume} {7}},\ \bibinfo
  {pages} {149} (\bibinfo {year} {1977})}\BibitemShut {NoStop}%
\bibitem [{\citenamefont {Kayser}(1981)}]{Kayser:1981ye}%
  \BibitemOpen
  \bibfield  {author} {\bibinfo {author} {\bibfnamefont {B.}~\bibnamefont
  {Kayser}},\ }\href {\doibase 10.1103/PhysRevD.24.110} {\bibfield  {journal}
  {\bibinfo  {journal} {Phys. Rev.}\ }\textbf {\bibinfo {volume} {D24}},\
  \bibinfo {pages} {110} (\bibinfo {year} {1981})}\BibitemShut {NoStop}%
\bibitem [{\citenamefont {Rich}(1993)}]{Rich:1993wu}%
  \BibitemOpen
  \bibfield  {author} {\bibinfo {author} {\bibfnamefont {J.}~\bibnamefont
  {Rich}},\ }\href {\doibase 10.1103/PhysRevD.48.4318} {\bibfield  {journal}
  {\bibinfo  {journal} {Phys. Rev.}\ }\textbf {\bibinfo {volume} {D48}},\
  \bibinfo {pages} {4318} (\bibinfo {year} {1993})}\BibitemShut {NoStop}%
\bibitem [{\citenamefont {Giunti}\ \emph {et~al.}(1993)\citenamefont {Giunti},
  \citenamefont {Kim}, \citenamefont {Lee},\ and\ \citenamefont
  {Lee}}]{Giunti:1993se}%
  \BibitemOpen
  \bibfield  {author} {\bibinfo {author} {\bibfnamefont {C.}~\bibnamefont
  {Giunti}}, \bibinfo {author} {\bibfnamefont {C.}~\bibnamefont {Kim}},
  \bibinfo {author} {\bibfnamefont {J.}~\bibnamefont {Lee}}, \ and\ \bibinfo
  {author} {\bibfnamefont {U.}~\bibnamefont {Lee}},\ }\href {\doibase
  10.1103/PhysRevD.48.4310} {\bibfield  {journal} {\bibinfo  {journal}
  {Phys.Rev.}\ }\textbf {\bibinfo {volume} {D48}},\ \bibinfo {pages} {4310}
  (\bibinfo {year} {1993})},\ \Eprint {http://arxiv.org/abs/hep-ph/9305276}
  {arXiv:hep-ph/9305276 [hep-ph]} \BibitemShut {NoStop}%
\bibitem [{\citenamefont {Akhmedov}\ and\ \citenamefont
  {Kopp}(2010)}]{Akhmedov:2010ms}%
  \BibitemOpen
  \bibfield  {author} {\bibinfo {author} {\bibfnamefont {E.~K.}\ \bibnamefont
  {Akhmedov}}\ and\ \bibinfo {author} {\bibfnamefont {J.}~\bibnamefont
  {Kopp}},\ }\href {\doibase 10.1007/JHEP04(2010)008} {\bibfield  {journal}
  {\bibinfo  {journal} {JHEP}\ }\textbf {\bibinfo {volume} {1004}},\ \bibinfo
  {pages} {008} (\bibinfo {year} {2010})},\ \Eprint
  {http://arxiv.org/abs/1001.4815} {arXiv:1001.4815 [hep-ph]} \BibitemShut
  {NoStop}%
\bibitem [{\citenamefont {Davis}\ \emph {et~al.}(1968)\citenamefont {Davis},
  \citenamefont {Harmer},\ and\ \citenamefont {Hoffman}}]{Davis:1968cp}%
  \BibitemOpen
  \bibfield  {author} {\bibinfo {author} {\bibfnamefont {R.}~\bibnamefont
  {Davis}, \bibfnamefont {Jr.}}, \bibinfo {author} {\bibfnamefont {D.~S.}\
  \bibnamefont {Harmer}}, \ and\ \bibinfo {author} {\bibfnamefont {K.~C.}\
  \bibnamefont {Hoffman}},\ }\href {\doibase 10.1103/PhysRevLett.20.1205}
  {\bibfield  {journal} {\bibinfo  {journal} {Phys. Rev. Lett.}\ }\textbf
  {\bibinfo {volume} {20}},\ \bibinfo {pages} {1205} (\bibinfo {year}
  {1968})}\BibitemShut {NoStop}%
\bibitem [{\citenamefont {Fukuda}\ \emph {et~al.}(1998)\citenamefont {Fukuda}
  \emph {et~al.}}]{Fukuda:1998mi}%
  \BibitemOpen
  \bibfield  {author} {\bibinfo {author} {\bibfnamefont {Y.}~\bibnamefont
  {Fukuda}} \emph {et~al.} (\bibinfo {collaboration} {Super-Kamiokande
  Collaboration}),\ }\href {\doibase 10.1103/PhysRevLett.81.1562} {\bibfield
  {journal} {\bibinfo  {journal} {Phys.Rev.Lett.}\ }\textbf {\bibinfo {volume}
  {81}},\ \bibinfo {pages} {1562} (\bibinfo {year} {1998})},\ \Eprint
  {http://arxiv.org/abs/hep-ex/9807003} {arXiv:hep-ex/9807003 [hep-ex]}
  \BibitemShut {NoStop}%
\bibitem [{\citenamefont {Ahmad}\ \emph {et~al.}(2002)\citenamefont {Ahmad}
  \emph {et~al.}}]{Ahmad:2002jz}%
  \BibitemOpen
  \bibfield  {author} {\bibinfo {author} {\bibfnamefont {Q.}~\bibnamefont
  {Ahmad}} \emph {et~al.} (\bibinfo {collaboration} {SNO Collaboration}),\
  }\href {\doibase 10.1103/PhysRevLett.89.011301} {\bibfield  {journal}
  {\bibinfo  {journal} {Phys.Rev.Lett.}\ }\textbf {\bibinfo {volume} {89}},\
  \bibinfo {pages} {011301} (\bibinfo {year} {2002})},\ \Eprint
  {http://arxiv.org/abs/nucl-ex/0204008} {arXiv:nucl-ex/0204008 [nucl-ex]}
  \BibitemShut {NoStop}%
\bibitem [{\citenamefont {Eguchi}\ \emph {et~al.}(2003)\citenamefont {Eguchi}
  \emph {et~al.}}]{Eguchi:2002dm}%
  \BibitemOpen
  \bibfield  {author} {\bibinfo {author} {\bibfnamefont {K.}~\bibnamefont
  {Eguchi}} \emph {et~al.} (\bibinfo {collaboration} {KamLAND Collaboration}),\
  }\href {\doibase 10.1103/PhysRevLett.90.021802} {\bibfield  {journal}
  {\bibinfo  {journal} {Phys.Rev.Lett.}\ }\textbf {\bibinfo {volume} {90}},\
  \bibinfo {pages} {021802} (\bibinfo {year} {2003})},\ \Eprint
  {http://arxiv.org/abs/hep-ex/0212021} {arXiv:hep-ex/0212021 [hep-ex]}
  \BibitemShut {NoStop}%
\bibitem [{\citenamefont {Abe}\ \emph {et~al.}(2012)\citenamefont {Abe} \emph
  {et~al.}}]{Abe:2012gx}%
  \BibitemOpen
  \bibfield  {author} {\bibinfo {author} {\bibfnamefont {K.}~\bibnamefont
  {Abe}} \emph {et~al.} (\bibinfo {collaboration} {T2K Collaboration}),\
  }\href@noop {} {\  (\bibinfo {year} {2012})},\ \Eprint
  {http://arxiv.org/abs/1201.1386} {arXiv:1201.1386 [hep-ex]} \BibitemShut
  {NoStop}%
\bibitem [{\citenamefont {Adamson}\ \emph {et~al.}(2012)\citenamefont {Adamson}
  \emph {et~al.}}]{Adamson:2012rm}%
  \BibitemOpen
  \bibfield  {author} {\bibinfo {author} {\bibfnamefont {P.}~\bibnamefont
  {Adamson}} \emph {et~al.} (\bibinfo {collaboration} {MINOS Collaboration}),\
  }\href@noop {} {\  (\bibinfo {year} {2012})},\ \Eprint
  {http://arxiv.org/abs/1202.2772} {arXiv:1202.2772 [hep-ex]} \BibitemShut
  {NoStop}%
\bibitem [{\citenamefont {Feynman}\ and\ \citenamefont
  {Gell-Mann}(1958)}]{Feynman:1958ty}%
  \BibitemOpen
  \bibfield  {author} {\bibinfo {author} {\bibfnamefont {R.~P.}\ \bibnamefont
  {Feynman}}\ and\ \bibinfo {author} {\bibfnamefont {M.}~\bibnamefont
  {Gell-Mann}},\ }\href {\doibase 10.1103/PhysRev.109.193} {\bibfield
  {journal} {\bibinfo  {journal} {Phys. Rev.}\ }\textbf {\bibinfo {volume}
  {109}},\ \bibinfo {pages} {193} (\bibinfo {year} {1958})}\BibitemShut
  {NoStop}%
\bibitem [{\citenamefont {Gell-Mann}\ and\ \citenamefont
  {Levy}(1960)}]{GellMann:1960np}%
  \BibitemOpen
  \bibfield  {author} {\bibinfo {author} {\bibfnamefont {M.}~\bibnamefont
  {Gell-Mann}}\ and\ \bibinfo {author} {\bibfnamefont {M.}~\bibnamefont
  {Levy}},\ }\href {\doibase 10.1007/BF02859738} {\bibfield  {journal}
  {\bibinfo  {journal} {Nuovo Cim.}\ }\textbf {\bibinfo {volume} {16}},\
  \bibinfo {pages} {705} (\bibinfo {year} {1960})}\BibitemShut {NoStop}%
\bibitem [{\citenamefont {Weinberg}(1991)}]{Weinberg:1991um}%
  \BibitemOpen
  \bibfield  {author} {\bibinfo {author} {\bibfnamefont {S.}~\bibnamefont
  {Weinberg}},\ }\href {\doibase 10.1016/0550-3213(91)90231-L} {\bibfield
  {journal} {\bibinfo  {journal} {Nucl.Phys.}\ }\textbf {\bibinfo {volume}
  {B363}},\ \bibinfo {pages} {3} (\bibinfo {year} {1991})}\BibitemShut
  {NoStop}%
\bibitem [{\citenamefont {Nakamura}\ \emph {et~al.}(2010)\citenamefont
  {Nakamura} \emph {et~al.}}]{Nakamura:2010zzi}%
  \BibitemOpen
  \bibfield  {author} {\bibinfo {author} {\bibfnamefont {K.}~\bibnamefont
  {Nakamura}} \emph {et~al.} (\bibinfo {collaboration} {Particle Data Group}),\
  }\href {\doibase 10.1088/0954-3899/37/7A/075021} {\bibfield  {journal}
  {\bibinfo  {journal} {J.Phys.G}\ }\textbf {\bibinfo {volume} {G37}},\
  \bibinfo {pages} {075021} (\bibinfo {year} {2010})}\BibitemShut {NoStop}%
\bibitem [{\citenamefont {G{\l}azek}(1987)}]{Glazek:1987cd}%
  \BibitemOpen
  \bibfield  {author} {\bibinfo {author} {\bibfnamefont {S.~D.}\ \bibnamefont
  {G{\l}azek}},\ }\href@noop {} {\bibfield  {journal} {\bibinfo  {journal}
  {Acta Phys. Pol.}\ }\textbf {\bibinfo {volume} {B18}},\ \bibinfo {pages} {85}
  (\bibinfo {year} {1987})}\BibitemShut {NoStop}%
\end{thebibliography}%

\end{document}